\documentclass[prb,twocolumn,showpacs,byrevtex,showkeys]{revtex4}

\usepackage{graphics}
\usepackage{amsmath}

\newcommand{\ie}{{\it i.e.}}
\newcommand{\eg}{{\it e.g.}}

\newcommand{\lhs}{l.h.s.}
\newcommand{\rhs}{r.h.s.}

\newcommand{\YBCO}{YBa$_2$Cu$_3$O$_7$}

\newcommand{\sgn}{\mathop{\mathrm{sgn}}\nolimits}

\newcommand{\am}{\mathop{\mathrm{am}}\nolimits}
\newcommand{\dn}{\mathop{\mathrm{dn}}\nolimits}
\newcommand{\F} {\mathop{\mathrm{F}}\nolimits}
\newcommand{\atanh}{\mathop{\mathrm{atanh}}\nolimits}

\newcommand{\state}[1]{$\stat{#1}$}

\newcommand{\stat}[1]{%
  \begingroup
  \def\delimiter##1##2##3##4##5##6##7##8{##10##3##4##5}%
  \mathcode`\0=\downarrow \mathcode`\1=\uparrow
  \mathcode`\d=\downarrow \mathcode`\u=\uparrow
  \mathcode`\D=\Downarrow \mathcode`\U=\Uparrow
  \mathcode`\-=\circ
  #1\relax
  \endgroup
}

\begin{document}

\title{
  Analytical analysis of ground states on 0-$\pi$ long Josephson junctions.
}

\author{A.~Zenchuk}
\email{zenchuk@itp.ac.ru}
\affiliation{Center of Nonlinear Studies of L.D.Landau Institute
for Theoretical Physics
(International Institute of Nonlinear Science)
Kosygina 2, Moscow, 119334,Russia
}

\author{E.~Goldobin}
\email{gold@uni-tuebingen.de}
\homepage{http://www.geocities.com/e_goldobin}
\affiliation{
  Institut f\"ur Mikro- und Nanoelektronische Systeme,
  Universitat Karlsruhe (TH),
  Hertzstrasse 16,
  D-76187 Karlsruhe, Germany
}
\affiliation{
  Physikalisches Institut - Experimentalphysik II,
  Universit\"at T\"ubingen,
  Auf der Morgenstelle 14,
  D-72076 T\"ubingen, Germany
}

\pacs{
  74.50.+r,   
  85.25.Cp    
  74.20.Rp    
}

\keywords{
  Long Josephson junction, sine-Gordon,
  half-integer flux quantum, semifluxon,
  0-pi-junction
}


\date{\today}

\begin{abstract}

  We investigate analytically a long Josephson 0-$\pi$-junction with several $0$ and $\pi$ facets which are comparable to the Josephson penetration length $\lambda_J$. Such junctions can be fabricated exploiting (a) the $d$-wave order parameter symmetry of cuprate superconductors; (b) the spacial oscillations of the order parameter in superconductor-insulator-ferromagnet-superconductor structures with different thicknesses of ferromagnetic layer to produce 0 or $\pi$ coupling or (c) the structure of the corresponding sine-Gordon equations and substituting the phase $\pi$-discontinuities by the artificial current injectors.
  We investigate analytically the possible ground states in such a system and show that there is a critical facet length $a_c$, which separates the states with half-integer flux quanta (semifluxons) from the trivial ``flat phase state'' without magnetic flux. We analyze different branches of the bifurcation diagram, derive a system of transcendental equations which can be effectively solved to find the crossover distance $a_c$ (bifurcation point) and present the solutions for different number of facets and the edge facets length. We show that the edge facets may drastically affect the state of the system.

\end{abstract}

\maketitle

\section{Introduction}
\label{Sec:Intro}

Due to the recent progress in technology it is now possible to fabricate different types of $\pi$ Josephson junctions (JJs): high-$T_c$ tri-crystal grain boundary JJs\cite{Tsuei:Review}, \YBCO-Nb zigzag ramp JJ\cite{Smilde:ZigzagPRL}, Superconductor-Ferromagnet-Superconductor (SFS)\cite{Ryazanov:2001:SFS-PiJJ,Ryazanov:2002:SFS-PiArray} or Superconductor-Insulator-Ferromagnet-Superconductor (SIFS)\cite{Kontos:2002:SIFS-PiJJ}. $\pi$-junctions are very promising elements for Josephson electronics. It was already suggested that they can be used in analog\cite{Goldobin:SFFO} and digital\cite{Gerritsma:RSFQwoBias,Terzioglu:1998:CJJ-Logic} circuits in classical regime and for implementation of qubits\cite{Ioffe:1999:sds-waveQubit}.


In this paper, we focus on {\em long} Josephson junctions (LJJ) consisting of several 0 and $\pi$ parts (facets). We will call such junctions 0-$\pi$-LJJs. LJJs consisting of very short (and random) 0 and $\pi$ facets, which are naturally formed in 45$^\circ$ high-$T_c$ grain boundaries, were studied by R.~Mints and coauthors in a series of works (see Ref.~\onlinecite{Mints:2002:SplinteredVortices@GB} and refs. therein). We are more interested in facets with the length $a$ comparable to the Josephson penetration depth $\lambda_J$ like in artificially prepared structures. These are the sizes which will be used in potential devices based on fractional vortex dynamics, both in classical and in quantum ones\cite{Kato:1997:QuTunnel0pi0JJ,Stefanakis:ZFS/2}.

It was found
\cite{Xu:SF-shape,Goldobin:LJJwPiPts} that at the point where 0 and $\pi$ facets join, a new type of non-linear excitation may appear. This new non-linear solution of the properly modified sine-Gordon equation looks like a vortex and contains one half of the flux quantum and therefore is called ``semifluxon''. The semifluxon (SF) is always pinned at the joining point between $0$ and $\pi$ facets. The presence of SFs was demonstrated experimentally\cite{Hilgenkamp:zigzag:SF} by scanning SQUID microscopy on \YBCO-Nb ramp zigzag LJJs in the \emph{long facet limit}, \ie, when the length of the facets $a\gg \lambda_J$. SFs were also experimentally observed in the tri-crystal grain boundary LJJs
\cite{Kirtley:SF:HTSGB,Kirtley:SF:T-dep,Sugimoto:TriCrystal:SF}


In this work we study analytically the ground states of 0-$\pi$-LJJ with arbitrary number of alternating 0 and $\pi$ facets and the effect of edge facets on the type of the ground state. As it was shown earlier for some particular cases, there is a critical facet length $a_c$ which separates the domains with the two most natural lowest energy configurations: the flat phase state and antiferromagnetically (AFM) ordered array of semifluxons.

The joining points between 0 and $\pi$ facets we will call a phase discontinuity points since, \eg, in \YBCO-Nb zigzag LJJs, the Josephson phase $\phi(x)$ is $\pi$-discontinuous at these points. In the other types of junctions, \eg, SFS and SIFS, the Josephson phase is continuous, but one can finally arrive to the same equations making a proper substitution of variables: $\phi(x,t)=\mu(x,t)+\theta(x)$ \cite{Goldobin:LJJwPiPts}. Following Ref.~\onlinecite{Goldobin:LJJwPiPts} and regardless of the LJJ type we will denote discontinuous phase as $\phi(x)$, while continuous (magnetic) component of the phase as $\mu(x)$.

In section \ref{Sec:Model} we introduce the model and represent general solution $\mu(x)$ for arbitrary distribution of $\pi$-discontinuity points and present several examples. In section \ref{Sec:ac}, we calculate the crossover distances $a_c$ (corresponding to ``AFM ordered SF chain''--``flat phase state'' transition) for $N$ equidistantly distributed $\pi$-discontinuity points with the arbitrary length $b$ of edge facets. Finally
section \ref{Sec:Conclusion} summarizes our results.

\section{Model and general stationary solution}
\label{Sec:Model}

We consider finite length Josephson junction with $N$ "$0$-$\pi$" conjunction points ($\pi$-discontinuity points). Let the coordinates of these points be $x_j$, $j=1,\dots,N$, and the coordinates of the two ends of LJJ be $x_0$ and $x_{N+1}$. We will write and solve all equations in terms of the magnetic component of the phase $\mu(x)$ which is a continuous function\cite{Goldobin:LJJwPiPts}. Let $\mu^j(x)$, $j=0,\dots,N $, be the piece of $\mu(x)$, inside of $j$-th interval $x_{j}\le x \le x_{j+1}$. Thus in total we have $N+1$ intervals enumerated by $j=0,1,\dots,N$. The $\mu^{2 n},\;\;n=0,1, \dots$ correspond to "0" intervals, and $\mu^{2 n+1},\;\;n=0,1, \dots$ correspond to "$\pi$" intervals.
While we use superscript $j$ to denote a piece of the function $\mu(x)$ at $j$-th interval, we use the subscript $j$ to denote the value of the function
$\mu(x)$ at $x=x_j$, \ie, $\mu_j=\mu(x_j)$, $j=0,\dots,N+1$. Since we will look for a continuous solution $\mu(x)$, $\mu_j$ are uniquely defined.

In these notations the sin-Gordon equation reads\cite{Goldobin:LJJwPiPts}:
\begin{eqnarray}
  \mu^j_{xx}-\mu^j_{tt}=(-1)^j \sin \mu^j.
  \label{Eq:sG}
\end{eqnarray}
Later on, we will need the above time dependent equation for analysis of stability of some solutions. At this moment we write only the stationary version of Eq.~(\ref{Eq:sG}) which has the form
\begin{eqnarray}
  \mu^j_{xx}=(-1)^j \sin \mu^j.
  \label{Eq:FP}
\end{eqnarray}
It may be integrated once:
\begin{eqnarray}\label{der}
  \left( \mu^j_x \right)^2 = C_j -  2 (-1)^j  \cos \mu^j
  ,\quad j=0,1,\dots,N,
\end{eqnarray}
where  $C_j$ are integration constants for $j$-th interval.
We look for solutions with arbitrary, but equal boundary conditions
$\mu_x(x_0)=\mu_x(x_{N+1})$.
Since $\mu_x$ is the magnetic field\cite{Goldobin:LJJwPiPts}, we may consider either uniform field or non-uniform one. In the first case the boundary conditions at both ends should be equal as written above. For non-uniform field in addition to having $\mu_x(x_0) \ne \mu_x(x_{N+1})$, we have to include the $h_x(x)$ term
\cite{Goldobin:LJJwPiPts} in the Eqs.~(\ref{Eq:sG}) and (\ref{Eq:FP}). For the sake of simplicity in this paper we will consider only uniform fields.

End points $x_0$ and $x_{N+1}$ may be either finite or infinite. Infinitely long JJ is considered as limit of finite JJ. First we construct solutions for finite length LJJ.

\subsection{General stationary solution for finite length JJ}

We require that the derivative $\mu_x$ is continuous at $x=x_j$,
i.e. $\mu^j_x(x_{j})=\mu^{j+1}_x(x_{j})$, $j=1,\dots,N$. Then
\begin{eqnarray}\label{C}
  C_0 &=& \mu^0_x(x_0)^2 + 2 \cos \mu_0
  ,\label{C0}\\\nonumber
  C_j &=& C_{j-1} + 4 (-1)^j \cos \mu_{j}=
  \\\nonumber
  &=& \mu^0_x(x_0)^2+ 2 \cos \mu_0 + 4 \sum_{i=1}^{j} (-1)^{i}
  \cos \mu_i;\\\nonumber &&\quad j=1,\dots N
  ,\label{Cj}\\\nonumber
  C_{N}&=&\mu^{N}_x(x_{N+1})^2+ 2 (-1)^{N} \cos \mu_{N+1}
  \label{CN}
\end{eqnarray}
The above system involves two different expressions for $C_{N}$,
which  produce  the first relation among $\mu_j$:
\begin{eqnarray}\label{first}
 2 \cos \mu_0 +
  4 \sum_{i=1}^{N} (-1)^{i}\cos \mu_i=
  2 (-1)^{N} \cos \mu_{N+1}
  .\label{mu1}
\end{eqnarray}

After the integration of (\ref{der})
one gets inside of the $j$-th interval:
%
\begin{eqnarray}
  x-x^*_j
  = \pm \int\limits_{2 \pi n_{j}+\sigma_j\pi}^\mu
    \frac{d\nu }{\sqrt{C_{j}+2\cos(\nu+\sigma_j \pi)}}
  \nonumber\\
  = \alpha_j \int\limits_0^{(\mu-\sigma_j \pi-2\pi n_j)/2}
    \frac{d\nu}{\sqrt{1-\alpha_j^2\sin^2 \nu}}
  \nonumber\\
  = \alpha_j \F[(\mu-\sigma_j \pi-2 \pi n_j)/2,\alpha_j^2]
  ,\label{x0}\\
  \alpha_j=\pm \frac{2}{\sqrt{C_j+2}}.
\end{eqnarray}
where $n_j$ are some integers which may be different for each
particular interval; $\F(x,m)$ is the Elliptic Integral of the First Kind; the lower limit of integration in the first integral is convenient for representation of the final result in terms of elliptic functions. To write such limits of integration we use $x^*_{n}\neq x_n$ in the above equations. The integration constants $x^*_j$ can be expressed in terms of $\mu_j$ ($j=0,\dots,N$):
\begin{eqnarray}
  x^*_{j}=x_{j} -\alpha_j \F[(\mu_{j}-\sigma_j\pi-2\pi n_j)/2,\alpha_j^2]
  ,\label{xstar}
\end{eqnarray}
where the sign in $\alpha_j$ is taken from the condition that $x$
increases when one goes along the junction. It is altering
in each extremum of the function $\mu(x)$; the variable $\sigma$ is such that $\sigma_{2n}=1$, $\sigma_{2n+1}=0$, $n=0,1,\dots,N$.
Functions $\mu^j(x)$ can be expressed in terms of  elliptic functions from the Eq.~(\ref{x0}):
\begin{eqnarray}
  \mu^j=2 \pi n_j+ \sigma_j \pi
  + 2\am\left[\frac{x-x^*_j}{\alpha_j},\alpha_j^2\right]
  ,\label{mu}
\end{eqnarray}
where $\am(x,m)$ is the Jacobi amplitude. The values $\mu_j$, $j=0,\dots,N+1$, are the solutions of the following system of
$N+1$ equations ($j=0,\dots,N$)
\begin{eqnarray}
  \mu_{j+1} =2 \pi n_j +\sigma_j \pi
  + 2\am\left[\frac{x_{j+1}-x^*_j}{\alpha_j},\alpha_j^2\right]
  ,\label{muj}
\end{eqnarray}
and Eq.~(\ref{mu1}).

Below we will need  extremum values  of $\mu^j$ (if any):
\begin{eqnarray}\label{A}
  \mu^j_\mathrm{ex} =\pm \arccos\left(\frac{(-1)^j C_j}{2}\right)  + 2 \pi n_j,
\end{eqnarray}

Remember, that the $\mu^j(x)$ can be either the piece of monotonic or periodic function depending on $C_j$, namely:
\begin{eqnarray}
  |C_j| < 2&:&\mu^j(x) \mbox{ piece of periodic function}
  \label{CC}\\
  |C_j|\ge2&:&\mu^j(x) \mbox{ piece of monotonic function}
  \label{CC2}
\end{eqnarray}

From the above one can get the following restriction on the
possible values of the parameters $n_j$:
if $\mu^j_x(x_j)>0$, then $n_{j+1}\ge n_j$;
if $\mu^j_x(x_j)<0$, then $n_{j+1}\le n_j$.

Thus Eq.~(\ref{mu}) together with Eqs.~(\ref{xstar}) and (\ref{muj}) represents general formulae for all possible ground states in 0-$\pi$-LJJ of finite length with arbitrary number of discontinuity points and uniform magnetic field. Each particular ground states is characterized by the parameters $n_j$ and $\mu_j$. The later are solutions of the system (\ref{muj}). We emphasize, that this system may be either solvable or not, which depends on values $N$ and positions of discontinuity points $x_j$.

\subsection{General stationary solution for infinitely long JJ}

In this section we give some remarks regarding stationary solutions for infinitely long JJ. The
main difference between finite and infinite length is related to the edge facets. Let us consider the limit $x_0\to -\infty$, $x_{N+1}\to \infty$.
We admit the following zero field boundary conditions:
\begin{eqnarray}\label{bound_inf}
  \lim_{x\to\pm\infty} \mu_x =\lim_{x\to -\infty} \mu= 0
  ,\nonumber\\
  \lim_{x\to \infty} \mu  = 2 \pi n_{N} + (1-\sigma_{N}) \pi
\end{eqnarray}
Equations (\ref{C0})--(\ref{CN}) now have the form:
\begin{eqnarray}\label{Cinf}
  C_0=C_{N}=2
  ,\\\nonumber
  C_j=C_{j-1} + 4 (-1)^j \cos \mu_{j-1}=
  \\\nonumber
  2 + 4 \sum_{i=1}^{j} (-1)^{i} \cos \mu_i;
  \quad j=1,\dots N.
\end{eqnarray}
Eq.~(\ref{mu1}) is reduced to the following one:
\begin{eqnarray}
\sum_{i=1}^{N} (-1)^i \cos \mu_i =0
\end{eqnarray}
Expression (\ref{A}) for $\mu^j_\mathrm{ex}$ as well as Eqs.~(\ref{xstar})--(\ref{muj}) stay the same for the inner intervals. But for the edge facets ($j=0,N$) Eqs.~(\ref{mu}) should be replaced with:
\begin{subequations}
  \begin{eqnarray}
    \mu^0(x) &=&
    4\arctan \left[ G_0\exp\left(
      x\,s_1\right) \right];\\
    \mu^{N}(x) &=& (1-\sigma_{N}) \pi 
    \nonumber\\
    &+& 4\arctan \left[ G_{N}\exp\left(
      x\,s_{N}\right) \right]+2 \pi n_{N},
  \end{eqnarray}
  \label{mu0Nc}
\end{subequations}
where parameters $s_1$ and $s_{N}$ may be either $1$ or $(-1)$ depending on boundary conditions at infinities (\ref{bound_inf}). They define whether the function is increasing or decreasing. Integration constants $G_0$ and $G_{N}$ are defined by the equations
\begin{subequations}
  \begin{eqnarray}
    \mu^0(x_1)&=& 4\arctan \left[ G_0\exp\left(
      x_1\,s_1\right) \right];\\
    \mu^{N}(x_{N})&=&(1-\sigma_{N}) \pi
    \nonumber\\
    &+& 4\arctan \left[ G_{N}\exp\left(
      x_{N}\,s_{N}\right) \right]+2 \pi n_{N}
  \end{eqnarray}
  \label{G0Nc}
\end{subequations}
%

\subsection{Examples of particular solutions}

The problem to construct solutions is reduced to solving the
system (\ref{mu1}) and (\ref{muj}). We
consider examples of two types of solutions: the so-called AFM state \state{udud}and the state \state{uudd} for LJJ with equidistant distribution of
discontinuity points $x_j$\cite{Goldobin:SF-ReArrange}, zero field boundary conditions
\begin{eqnarray}
  \mu_x(x_0)=\mu_x(x_{N+1}) = 0
\end{eqnarray}
and $n_j=0$ for all $j$. Thus $0<\mu<2 \pi$,
\begin{eqnarray}\label{An}
  \mu^j_\mathrm{ex} = \arccos\left(\frac{(-1)^j C_j}{2}\right) .
\end{eqnarray}
In this case  expression for
$\alpha_j$ in the Eqs.~(\ref{x0})--(\ref{CC2}) can be given in the following form:
\begin{eqnarray}
  \alpha_j= \frac{2\sgn
  \mu_x(x_j)}{\sqrt{C_j+2}},\;\;j=1,\dots,N,\;\;
  \alpha_0= -  \frac{2}{\sqrt{C_0+2}},
\end{eqnarray}
Following Ref.~\onlinecite{Goldobin:SF-ReArrange}, we consider equidistant distribution of discontinuity points $x_j=a(j-1)$, $j=1,\dots,N$, but with end points $x_0=-b$ and $x_{N+1}=a (N-1)+b$, where $b$ is the length of the edge facets which may be different from $a$. Due to the symmetry one has

\begin{subequations}
  \begin{eqnarray}
    &&\mbox{For even } N,\quad j=0,\dots,N/2:\nonumber\\
    &&\quad\mu_j=\phantom{\pi-}\mu_{N+1-j}
    ,\label{symm-even}\\
    &&\mbox{For odd } N,\quad j=0,\dots,(N-1)/2:\nonumber\\
    &&\quad\mu_j=\pi-\mu_{N+1-j},\quad\mu_{(N+1)/2}=\pi/2
    .\label{symm-odd}
  \end{eqnarray}
  \label{Eq:symm}
\end{subequations}

\begin{figure*}[!htb]
  \begin{center}
    \hfill
    \resizebox{75mm}{!}{\includegraphics{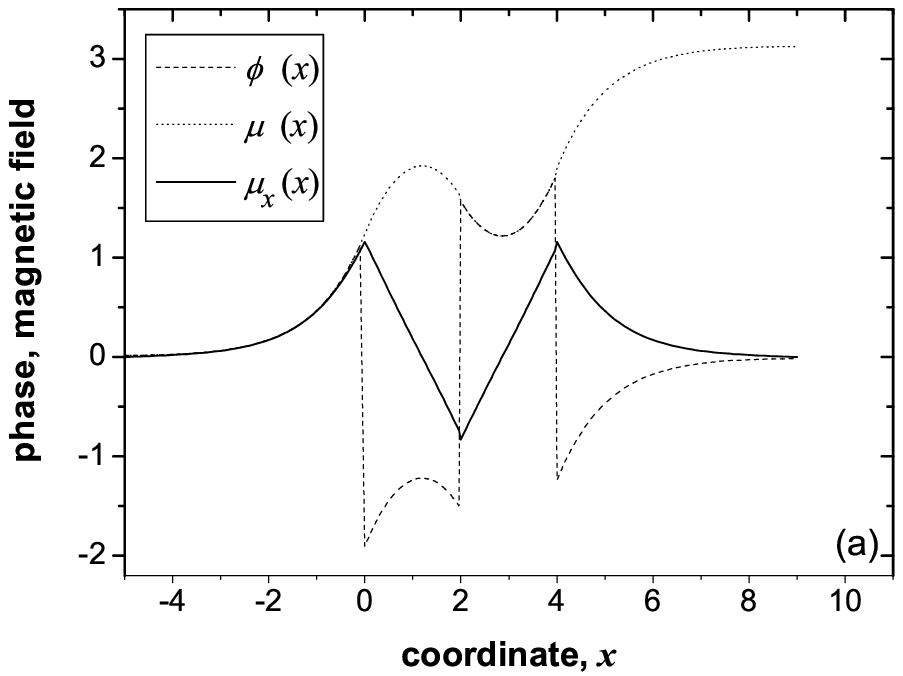}}
    \hfill
    \resizebox{75mm}{!}{\includegraphics{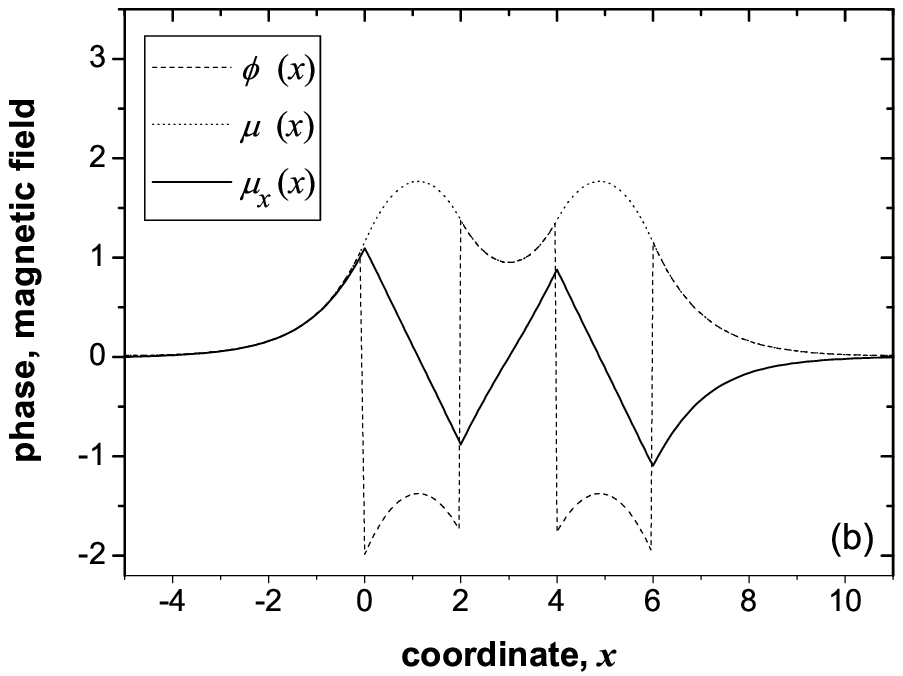}}
    \hfill
  \end{center}
  \caption{
    The state with AFM ordered SFs. Graphs of $\phi(x)$, $\mu(x)$ and $\mu_x(x)$ for $a=2$, $b=5$; (a) $N=3$, $\mu_0=0.0172$, $\mu_1=1.2371$ and (b) $N=4$, $\mu_0=0.016$, $\mu_1=1.1569$, $\mu_2=1.3775$.
  }
  \label{Fig:Plots:a2b5N34}
\end{figure*}

\paragraph{AFM state in finite length LJJ.}
For AFM state, $\mu^j$ should have extremum value inside its interval, which means that $\mu^j$ is a piece of periodic function (\ref{CC}), so that one needs to provide $ C_j<2 $ for all $j$. The plots of $\phi(x)$, $\mu(x)$, $\mu_x(x)$ for $a=2$, $b=5$ and two different values of $N$
($N=3$ and $N=4$) are shown in Fig.~\ref{Fig:Plots:a2b5N34}.

\begin{figure}[!htb]
  \begin{center}
    \resizebox{75mm}{!}{\includegraphics{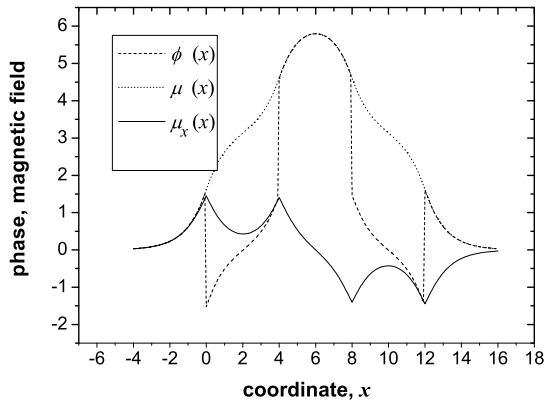}}
  \end{center}
  \caption{
    The state \protect\state{uudd}. Graphs of $\phi(x)$, $\mu(x)$ and $\mu_x(x)$ for
    infinitely long JJ with $a=4$ and  $N=4$; $\mu_1=1.6165$,
    $\mu_2=4.6086$.
  }
  \label{Fig:Plots:PSF}
\end{figure}

\paragraph{\protect\state{uudd} state in infinite LJJ}. Let us consider infinitely long JJ with $N=4$. For this state  $\mu^j$ should have extremum value inside of the middle interval: $C_1,C_3>2$; $C_0=C_4=2$, and $0<C_2<2$. Parameters in Eqs.~(\ref{mu0Nc}) and (\ref{G0Nc}) are defined as follows: $\sigma_{N}=1$, $s_1=1$, $s_2=-1$. The plots of $\phi(x)$, $\mu(x)$, $\mu_x(x)$ are shown in Fig.~\ref{Fig:Plots:PSF}

\section{Crossover distance}
\label{Sec:ac}

In this section we study the transition between flat phase state and AFM ordered SF state in 0-$\pi$-LJJ with equidistant distribution of $\pi$-discontinuity points $|x_{j+1}-x_j|=a$, $j=1,\dots,N-1$ expressing the length of the edge facets in terms of $a$: $b=\beta a$. For this case $0\le \mu\le \pi$.

In the next subsection \ref{Sec:FPS} we study the stability of the flat phase solution and derive the system of algebraic equations for calculation of $a_c$. In the subsection \ref{Sec:AFM-disappear} we study the existence of AFM solution and derive another system of equations defining $a_c$.

\subsection{Stability of flat phase state}
\label{Sec:FPS}

We study solutions of the time dependent Eq.~(\ref{Eq:sG}) which have small amplitude oscillations around the flat state $\mu_c$. As it follows from Eq.~(\ref{Eq:FP}), the value of phase $\mu(x)=\mu_c$ in the flat phase state can be either $0$ or $\pi$. Note that, for odd $N$ the symmetry conditions (\ref{symm-odd}) result in $\mu=\pi/2$ at the middle point $x=x_{N/2}$. Since $\mu(x)=const$ in the flat phase state, $\mu$ should be equal to $\pi/2$, but this is not a solution of Eq.~(\ref{Eq:sG}). Thus, we conclude that for odd $N$ the flat phase state cannot be realized, so we formally take $a_c=0$. Below we consider only even $N$.

Let us introduce a new function $\tilde \mu = \mu-\mu_c \ll 1$ for which equation (\ref{Eq:sG}) has one of the forms given below
\begin{subequations}
  \begin{eqnarray}
    \tilde\mu^j_{xx}-\tilde\mu^j_{tt}&=&(-1)^{j}  \tilde\mu,\quad\mu_c=0
    ,\label{Eq:mu_c=0}\\
    \tilde\mu^j_{xx}-\tilde\mu^j_{tt}&=&(-1)^{j+1}\tilde\mu,\quad\mu_c=\pi
    \label{Eq:mu_c=pi}
  \end{eqnarray}
  \label{Eq:tilde_mu}
\end{subequations}
for all $j=0,\dots,N+1$.

First, we study only the Eq.~(\ref{Eq:mu_c=0}) corresponding to $\mu_c=0$. To study the stability we look for the solution in the following form:
\begin{equation}
  \tilde \mu^j=e^{\sqrt{E}t} \nu^j,
  \label{Eq:E}
\end{equation}
where $E$ is considered around $E=0$, because this is the point where the stability of solution changes. Hereafter we use $-1<E<1$. Then Eq.~(\ref{Eq:mu_c=0}) gets the form
\begin{equation}
  \nu_{xx}^j= \left[ E+(-1)^j \right] \nu^j, \quad j=0,\dots,N+1.
  \label{Eq:nuj}
\end{equation}
There are two solutions of the correspondent characteristic equation:
$k=\pm k_1$ for even intervals and $k=\pm i k_2$ for odd intervals, where $k_1=\sqrt{1+E}$ and $k_2= \sqrt{1-E}$ are both real. The solutions of the Eqs.~(\ref{Eq:nuj}) can be represented  by the following system
\begin{subequations}
  \begin{eqnarray}
    \nu^0&=&A_0 \cosh(k_1 x +\beta_0)\\
    \nu^j&=&A_j \cosh[k_1 (x-a (j-1))+\beta_j],\;j=2,4,\dots,\\
    \nu^j&=&A_j \cos[k_2 (x-a (j-1))+\beta_j],\;j=1,3,\dots.
  \end{eqnarray}
  \label{Eq:small_ampl_sol}
\end{subequations}
The zero field boundary condition $\mu_x(-\beta a)=0$ gives the formula for $\beta_0$: $\beta_0=\beta a$. Due to the symmetry (\ref{symm-odd}), it is enough  to consider only half of the whole LJJ. We have  for the middle interval $\mu_{N/2}=\mu_{N/2+1}$, which gives expressions for $\beta_{N/2}$:  $\beta_{N/2}=-k_1 a/2$ for even  $N/2$, or $\beta_{N/2}=-k_2 a/2$ for odd $N/2$. Continuity of the functions $\mu$ imposes the following relations among parameters $A_j$ and $\beta_j$:
\begin{subequations}
  \begin{eqnarray}
    A_0 \cosh(\beta a) &=&  A_1 \cos(\beta_1)
    ,\\
    A_j\cos( k_2 a+\beta_j) &=& A_{j+1}\cosh(\beta_{j+1})
    ,\; j=1,3,\dots\\
    A_j\cosh( k_1 a+\beta_j) &=& A_{j+1}\cos(\beta_{j+1})
    ,\; j=2,4,\dots
  \end{eqnarray}
  \label{Eq:mu_cont}
\end{subequations}
To provide continuity of $\mu_x$, one needs to impose additional relations among the parameters $\beta_j$:
\begin{subequations}
  \begin{eqnarray}
    k_1\tanh(\beta a) &=&  -k_2\tan(\beta_1)
    ,\label{Eq:mu_x_cont1}\\
    k_2\tan( k_2 a+\beta_j) &=&  -k_1\tanh(\beta_{j+1})
    ,\quad j=1,3,\dots, \label{Eq:mu_x_cont2}\\
    k_1\tanh( k_1 a+\beta_j) &=& -k_2\tan(\beta_{j+1})
    ,\quad j=2,4,\dots, \label{Eq:mu_x_cont3}
  \end{eqnarray}
\end{subequations}
where $j< N/2$.
Equations (\ref{Eq:mu_cont}) define the amplitudes $A_j$, $j>0$, in terms of $A_0$ and $\beta_n$, $n=0,\dots,N/2$.
The Eq.~(\ref{Eq:mu_x_cont1}) establishes relation between $a$ and $E$, while Eqs.~(\ref{Eq:mu_x_cont2}) and (\ref{Eq:mu_x_cont3}) define parameters $\beta_j$.

The solution (\ref{Eq:E}) is stable if $E\le 0$.
We define crossover distance $a_c$ the distance for which $E=0$
and, consequently, $k_1=k_2=1$, \ie, the system of Eqs.~(\ref{Eq:mu_x_cont1})--(\ref{Eq:mu_x_cont3}) defining $a_c$ is
\begin{equation}\left\{
  \begin{array}{rcl}
    \tanh(\beta a_c)  &=&  -\tan (\beta_1),\\
    \tan (a_c+\beta_j)&=&  -\tanh(\beta_{j+1}),\;\;j=1,3,\dots\\
    \tanh(a_c+\beta_j)&=&  -\tan (\beta_{j+1}),\;\;j=2,4,\dots\\
    \beta_{N/2}     &=& -\frac{a_c}{2}.
  \end{array}\right.
  \label{Eq:a_c-instability1}
\end{equation}
Note that this system of equations is rather easy to solve consequently excluding $\beta_j$. At the end one gets a transcendental equation for $a_c$ [the first equation of (\ref{Eq:a_c-instability1})], with given parameter $\beta$ and function $\beta_1(a_c)$. It is clear, that $a_c=0$ and all $\beta_j=0$ is solution of the system (\ref{Eq:a_c-instability1}). Instead, we are interested to find the first non-zero solution $a_c$ of (\ref{Eq:a_c-instability1}).

Non-zero solution to this transcendental equation does not exists for any $\beta$. To derive the existence conditions we analyze the behavior of the function $\tan[-\beta_1(a)]$. From Eqs.~(\ref{Eq:a_c-instability1}) it is evident that $\tan[-\beta_1(a)]$ is monotonic function of $a$ with uniform concavity. This conclusion is direct consequence of the fact that both $\tan(x)$ and $\tanh(x)$  in the system (\ref{Eq:a_c-instability1}) are monotonic with uniform concavity inside of the continuity interval $0\le x \le \pi/2$. Thus, if $\tan[-\beta_1(a)]$
increases in some particular point, then it will be increasing inside the whole interval. Let us find the behavior of $\tan(-\beta_1(a))$ for $a\to 0$. All but first equations of the system  (\ref{Eq:a_c-instability1}) are easily resolvable for $\beta_j$, $j>1$ giving
$\beta_j\approx -a/2$.
Thus $\tan\left[ -\beta_1(a) \right]$ is an increasing function of $a$. Since $\tan(x)\to\infty$ for $x\to\pi/2$, this function is concaved up. Since $\tanh(\beta a)$ is concaved down, the two functions may intersect [the first of Eqs.~(\ref{Eq:a_c-instability1}) has nonzero solution], only if
$\tanh'(0)>\tan'(\beta_1(0))$ (the prime denotes a derivative with respect to $a$), \ie, if $\beta>1/2$. Using  Eq.~(\ref{Eq:mu_x_cont1}) we have
\begin{equation}
  \sqrt\frac{E+1}{1-E}=
  \frac{\tan(-\beta_1(a))}{\tanh(\beta a)}\le 1
  \label{Eq:noneq}
\end{equation}
and, consequently, $E\le 0$ inside of the interval $0\le a\le a_c$.

We have found that for any $\beta>1/2$ there is an interval $0< a <a_c$ where the flat phase state $\mu_c=0$ is stable. If  $\beta<1/2$, then $E>0$ and the flat phase state $\mu_c=0$ is unstable. Thus, $\beta=1/2$ is a threshold value, for which $E=0$ and $a_c=0$.

In a similar way, one can show that the flat phase state $\mu_c=\pi$, corresponding to Eq.~(\ref{Eq:mu_c=pi}), is stable inside the interval $0<a<a_c$, if $\beta<1/2$. The appropriate system defining $a_c$ is:
\begin{equation}\left\{
  \begin{array}{rcl}
    \tan (\beta a_c)  &=&  -\tanh(\beta_1),\\
    \tanh(a_c+\beta_j)&=&  -\tan (\beta_{j+1}),\;\;j=1,3,\dots\\
    \tan (a_c+\beta_j)&=&  -\tanh(\beta_{j+1}),\;\;j=2,4,\dots\\
    \beta_{N/2}     &=& -\frac{a_c}{2}.
  \end{array}\right.
  \label{Eq:a_c-instability2}
\end{equation}

The values of $a_c$ corresponding to the instability of the flat phase state are calculated using Eqs.~(\ref{Eq:a_c-instability1}) and (\ref{Eq:a_c-instability2}) for $\beta=\frac{1}{4},1,2,\infty$, and summarized in Tab.~\ref{Tab:ac-instability}.
One can check that the values in this table are in accordance with those obtained earlier by direct numerical simulation \cite{Goldobin:SF-ReArrange} for $\beta=1/2,1$. The result for $N=2$ and $\beta=\infty$ coincides with the one calculated earlier analytically\cite{Kato:1997:QuTunnel0pi0JJ,Kuklov:1995:Current0piLJJ}. We stress here that our present results are obtained for arbitrary edge facet length $b$ (arbitrary $\beta$) and have much higher accuracy (can be calculated much faster).

\begin{figure}[!htb]
  \centering\includegraphics{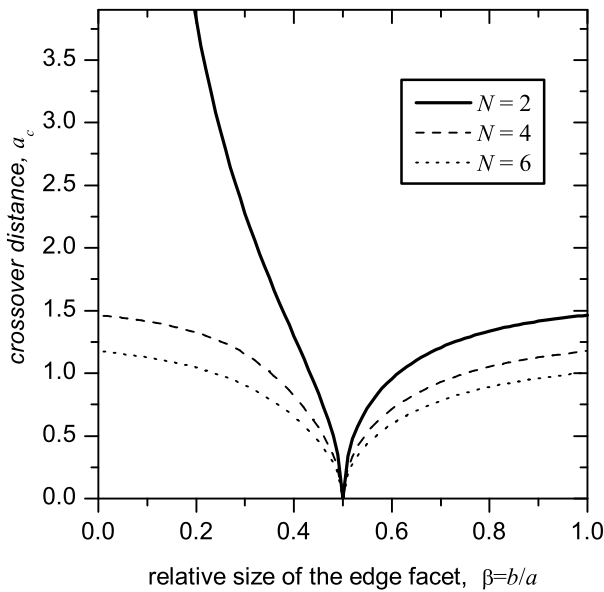}
  \caption{
    The dependences of crossover distances $a_c^{(2)}$, $a_c^{(4)}$ and $a_c^{(6)}$ on $\beta$. These curves can be obtained by solving either stability problem [Eqs.~(\ref{Eq:a_c-instability1}) for $b>a/2$ or Eqs.~(\ref{Eq:a_c-instability2}) for $b<a/2$] or
    AFM solution existence problem [Eqs.~(\protect\ref{Eq:ac-LargeBeta}) for $b>a/2$ and Eqs.~(\protect\ref{Eq:ac-SmallBeta}) for $b<a/2$].
  }
  \label{Fig:a_c(b)}
\end{figure}

The plot $a_c^{(N)}(\beta)$ for different $N$ can be seen in Fig.~\ref{Fig:a_c(b)}. As $\beta\to0$, the $a_c^{(2)}\to\infty$ while for $N>2$ the $a_c^{(N)}$  approaches some finite value. Also note the following natural property which can be seen in Fig.~\ref{Fig:a_c(b)}: $a_c^{(N)}(1)=a_c^{(N+2)}(0)$. The fact that $a_c^{(2)}\to\infty$ for $\beta\to0$ means that $0-\pi-0$ LJJ approaches the limit of all-$\pi$ LJJ, where vortex solutions are unstable and flat phase state wins.

\begin{table}
  \begin{equation*}
    \begin{array}{|r|c|c|c|c|}
      \hline
          & \multicolumn{4}{c|}{a_c} \cr
      \hline
      N & \beta=1/4 & \beta=1 & \beta=2 & \beta=\infty \cr
      \hline
      2 & 2.92771 & 1.4639 & 1.5670 & \pi/2\cr
      4 & 1.25461 & 1.1772 & 1.2989 & 1.3063\cr
      6 & 0.98327 & 1.0060 & 1.1245 & 1.1343\cr
      8 & 0.83438 & 0.8914 & 1.0032 & 1.0146\cr
      10& 0.73731 & 0.8082 & 0.9134 & 0.9259\cr
      \hline
    \end{array}
  \end{equation*}
  \caption{
    The values of $a_c^{(N)}$ for $\beta=1/4,\,1,\,2,\,\infty$
    corresponding to instability of flat phase state. (accuracy of calculation
    is $\pm 0.0001$)
 }
  \label{Tab:ac-instability}
\end{table}

\subsubsection{Behavior of instability point at large $N$}
\label{Sec:a_c@LargeN}

Using the systems (\ref{Eq:a_c-instability1}) and (\ref{Eq:a_c-instability2}) its possible to determine a ground state of a very large LJJ, \ie, when $N\to\infty$.

First, let us consider the case $\beta<1/2$. As we saw above, according to the system (\ref{Eq:a_c-instability2}) when $a_c\to0$, the values of $\beta_j \sim a_c$. We write $\beta_j=a_c\psi_j$, where $\psi_j\sim 1$. In this case we expand the equations of the system (\ref{Eq:a_c-instability2}) in Taylor series for small $a_c$ and discard all the terms smaller than $O(a_c^3)$:
\begin{equation}\left\{
  \begin{array}{rcl}
    \beta  &\approx& -\psi_1 + \frac{2}{3} a_c^2 \psi_1^3,\\
    \psi_j &\approx& -1-\psi_{j+1} - \frac{2}{3} a_c^2 \psi_{j+1}^3
    ,\quad j=1,3,\dots\\
    \psi_j &\approx& -1-\psi_{j+1} + \frac{2}{3} a_c^2 \psi_{j+1}^3
    ,\quad j=2,4,\dots\\
    \psi_{N/2}&=& -1/2.
  \end{array}\right.
  \label{Eq:a_c-instab2-expand}
\end{equation}
Substituting expressions for $\psi_j$ one by one starting with the last equation  and discarding all the terms smaller than $O(a_c^2)$ we result in the final expressions for $\psi_j$:
\begin{equation}
  \psi_j = -\frac{1}{2} 
  + (-1)^{j+1} \left( \frac{N}{2} - j \right)\frac{a_c^2}{12}
  .\label{Eq:AB-b0}
\end{equation}
Thus, the first Eq.\ of the system (\ref{Eq:a_c-instab2-expand}) now reads
\begin{equation}
  \beta \approx \frac{1}{2} - \frac{a_c^2}{24} N
  . \label{Eq:1st@LargeN-b0}
\end{equation}
Finally, the expression for $a_c$ is
\begin{equation}
   a_c \approx \sqrt{\left( \frac{1}{2}-\beta \right)\frac{24}{N}}
  . \label{Eq:a_c@LargeN-b0}
\end{equation}
It is valid for $N\gg 24(1/2-\beta)$.

Second, for {\it finite} $\beta>1/2$ the reasoning is similar, and we get
\begin{equation}
   a_c \approx \sqrt{\left( \beta-\frac{1}{2} \right)\frac{24}{N}}
  . \label{Eq:a_c@LargeN-b1}
\end{equation}
This approximation is valid for $N \gg 24(\beta - 1/2)\max(1,\beta^2)$.

In the case $\beta\to\infty$  ({\it infinitely long edge facets} ),  $\tanh(\beta a_c)\to 1$ and our approximation does not work. 
The reason is that $\beta_j$ does not vanishes with increase of $N$ in this case.

Alternative, more complex but also more exact, derivation of asymptotic
behavior of $a_c$ for $N\to\infty$ and any finite as well as
infinite $\beta$  is presented in appendix
\ref{Sec:a_c@LargeN_app}.


\subsection{Existence of AFM ordered solution}
\label{Sec:AFM-disappear}

In this section we apply the general formulas derived in the Sec.~\ref{Sec:Model} to calculate the crossover distance for AFM state in the LJJ with equidistant distribution of $\pi$-discontinuity points.

\begin{figure*}[!htb]
  \begin{center}
    \hfill
    \resizebox{!}{40mm}{\includegraphics{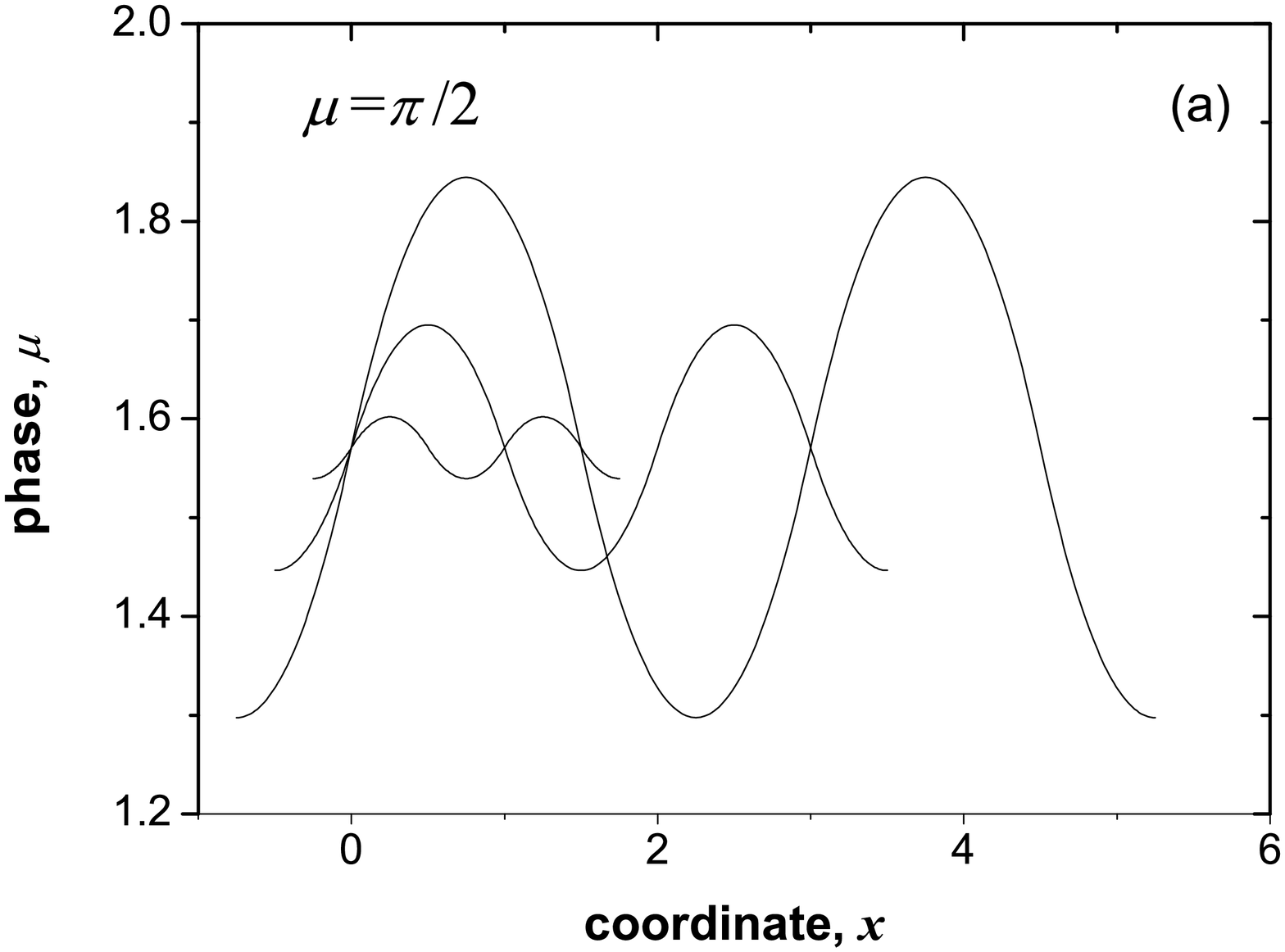}}
    \hfill
    \resizebox{!}{40mm}{\includegraphics{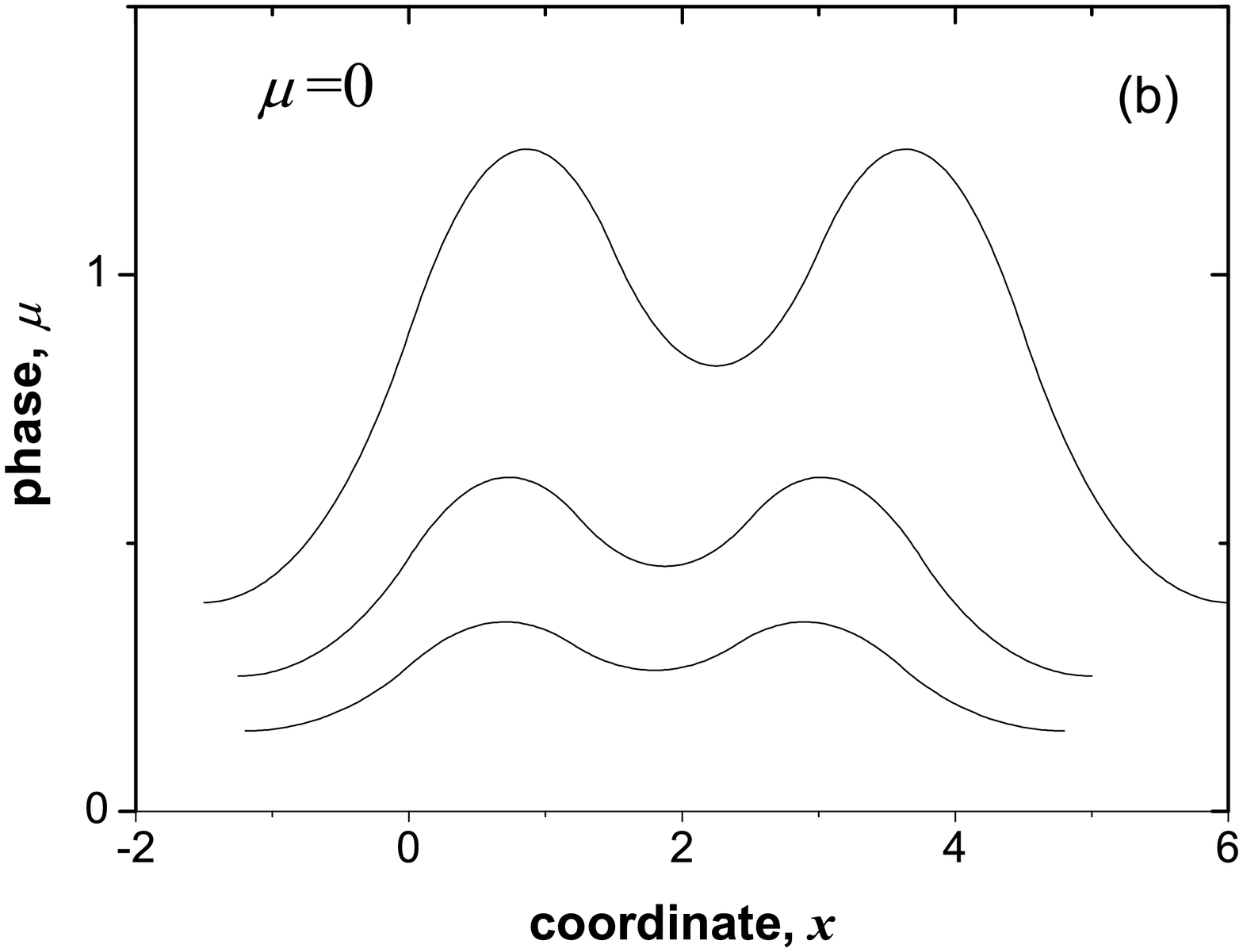}}
    \hfill
    \resizebox{!}{40mm}{\includegraphics{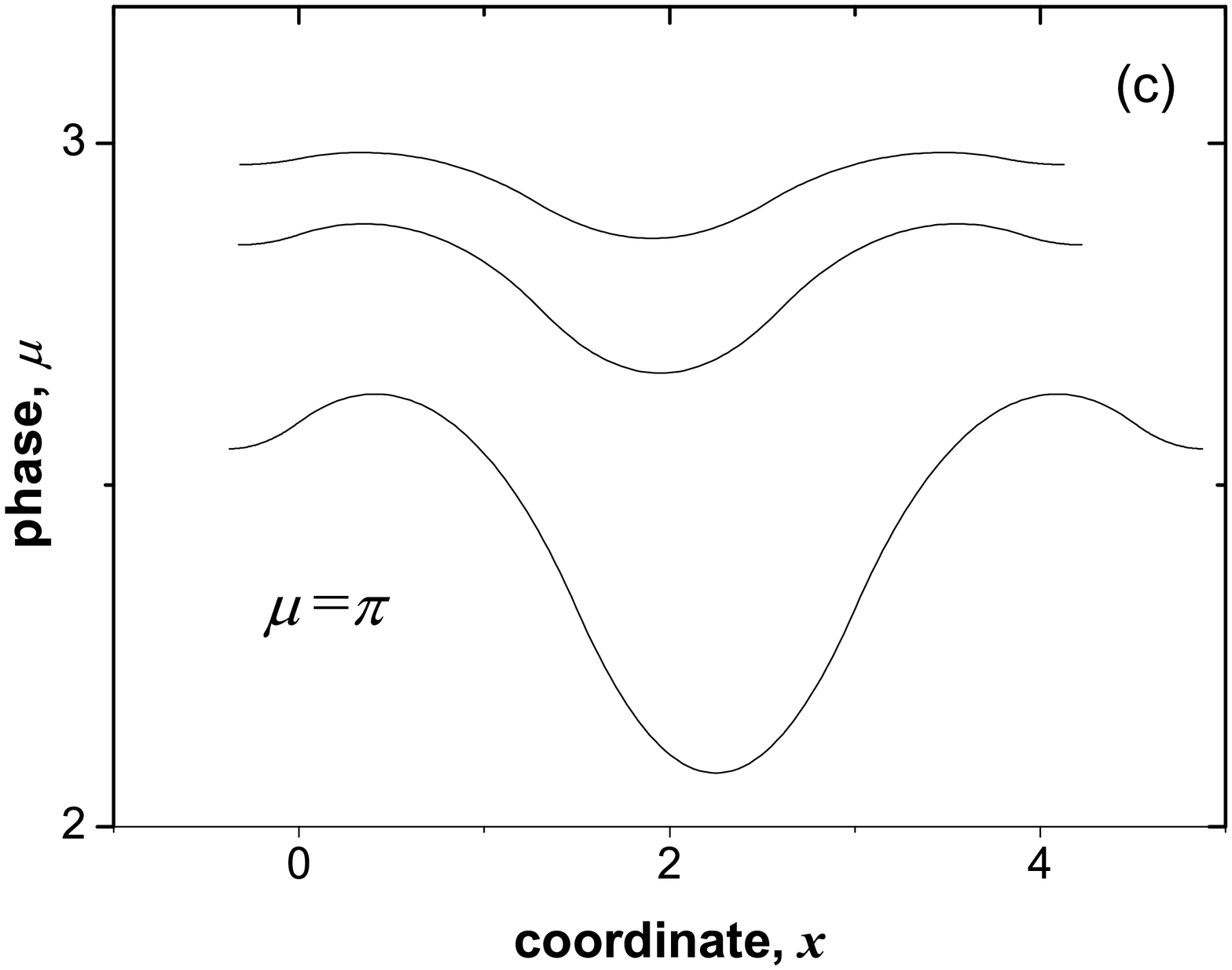}}
    \hfill
  \end{center}
  \caption{
    Graphs of $\mu(x)$, amplitude decreases
    with
    decrease of $a$: (a)
     $b=a/2, N=4$; $a= 3/2,1,1/2$, $\mu_c=\pi/2$; (b) $b=a,N=4$; $a= 3/2,1.25,1.2$, $\mu_c=0$; (c) $b=a/4,N=4$; $a= 3/2,1.3,1.27$, $\mu_c=\pi$
  }
  \label{Fig:L0=Lpi}
\end{figure*}

If one chooses a domain of parameters where AFM ordered SF chain is present, vary $a$ and plot the solutions $\mu(x)$, he may notice that the amplitude of spatial oscillations of $\mu(x)$ decreases as $a$ decreases, see Fig.~\ref{Fig:L0=Lpi}. It may happen, as we show below, that the amplitude of $\mu(x)$ vanishes at some $a=a_c$ which may be larger than zero. 

In this section we re-introduce the crossover distance $a_c$ as a distance at which the amplitude of oscillations of $\mu(x)$ vanishes, \ie, $\mu(x)\to\mu_c$ for $a\to a_c+0$. In fact the limiting value of the phase $\mu_c$ can have only three different values: $0,\;\pi/2,\;\pi$.
To prove this, we refer to the Eq.~(\ref{x0}) and write the set of
expressions for the distance $a=|x(\mu_j+1)-x(\mu_j)|$
($j=1,\dots,N-1$) between discontinuity points
\begin{widetext}
  \begin{eqnarray}
    a = (-1)^{j+1} \left(
      \int\limits_{\mu_{j}}^{\mu^j_\mathrm{ex}}
      \frac{d\nu}{\sqrt{C_j-2 (-1)^j\cos\nu}}
      - \int\limits^{\mu_{j+1}}_{\mu^j_\mathrm{ex}}
      \frac{d\nu}{\sqrt{C_j-2 (-1)^j\cos\nu}}
    \right)
    \label{a}
  \end{eqnarray}
\end{widetext}
$j=1,\dots,N$. For the left edge interval it reads:
\begin{eqnarray}
  \beta a = \int\limits^{\mu_{1}}_{\mu^0}
  \frac{d\nu}{\sqrt{C_0-2 \cos\nu}}
  ,\label{bl}\label{Eq:b}
\end{eqnarray}
Due to the symmetry conditions (\ref{Eq:symm}) we consider here only the left half of the junction. Note that in spite of index $j$, both sides of the   expression (\ref{a}) do not depend on the facet number $j$, since we consider equidistantly distributed discontinuity points. Consider the limit $a\to a_c+0$, which means
\begin{equation}
  \mu_j\approx \mu_c + \epsilon \tilde\mu_j
  , \label{Eq:mu-approx}
\end{equation}
where $\epsilon\to0$. The integration intervals are proportional to $\epsilon$. The denominator can be approximated using $\nu=\mu_c+\epsilon \tilde\nu$ as:
\[
  \sqrt{-2\epsilon K_s
  \sin\mu_c -\epsilon^2 K_c
  \cos\mu_c},
\]
where $K_s$ and $K_c$ do not depend neither on $\mu_c$ nor on $\epsilon$. If $\mu_c=0$ or $\mu_c=\pi$, the denominator in Eq.~(\ref{a}) is $\propto\epsilon$, which results in $a_c>0$. Otherwise the denominator is $\propto\sqrt{\epsilon}$, so that $a_c=0$.

For all other values of $\mu_c\ne0$ and $\mu_c\ne\pi$, the denominator in Eq.~(\ref{a}) is constant and integral vanishes, \ie, $a_c=0$. It is interesting that ``all other values of $\mu_c$'' essentially mean $\mu_c=\pi/2$, see  appendix \ref{Sec:mu_c=pi/2} for details. This also allows us to make a quick conclusion that $a_c=0$ for odd $N$. Indeed, for odd $N$, due to the symmetry conditions (\ref{symm-odd}) $\mu_{(N+1)/2}=\pi/2$ for any $a$. Therefore, $\mu(x)\to\mu_c=\pi/2$ when $a$ decreases. This automatically means that $a_c=0$.

Summarizing our findings, one can get the following possible values of $\mu_c$ depending on $a_c$ and $\beta$:
\begin{enumerate}
  \item $\mu_c=\pi/2$ for odd $N$, Fig.~\ref{Fig:L0=Lpi}(a); in this case $a_c=0$.
  \item $\mu_c=0$ for even $N$ and $\beta>1/2$, Fig.~\ref{Fig:L0=Lpi}(b); in this case $a_c>0$.
  \item $\mu_c=\pi$ for even $N$ and $\beta<1/2$, Fig.~\ref{Fig:L0=Lpi}(c); in this case $a_c>0$.
\end{enumerate}

Since, for odd $N$ the $a_c=0$ is already known, we calculate $a_c$ only for the last two cases.

Since for AFM state one has  the symmetry (\ref{symm-even}). So,
it is enough to consider $j=0,\dots N/2$ with condition
$\mu(a_{N/2})=\mu(a_{N/2-1})$.

\emph{Even $N$, $\beta>1/2$, $\mu_c=0$}. Using Eq.~(\ref{Eq:mu-approx}) with $\mu_c=0$ the Eqs.~(\ref{C}) and (\ref{An}) can be approximated as follows:
\begin{eqnarray} \label{Cappr}
  C_j \approx
  2 (-1)^j  +\epsilon^2 \Sigma_j
  ,\quad\label{Capr0}\\\label{mujappr}
  \mu^j_\mathrm{ex}\approx\epsilon \sqrt{(-1)^{j+1}\Sigma_j}=
  \epsilon\tilde\mu^j_\mathrm{ex}
  ,\label{Aapr0}
\end{eqnarray}
where we defined $\Sigma_j$ as
\begin{eqnarray}
  \Sigma_j=-2\sum_{i=1}^{j}(-1)^i \tilde\mu_i^2 - \tilde\mu_0^2
  .\label{Sigma_j0}
\end{eqnarray}
It follows from the Eqs.~(\ref{CC}) and (\ref{Capr0}) that
$\Sigma_j>0$ for odd $j$ and $\Sigma_j<0$ for even $j$.

In the limit $\epsilon\to0$ the Eqs.~(\ref{a}) and (\ref{bl}) become:
\begin{widetext}
  \begin{eqnarray}
    a_c^{(N)} &=& (-1)^{j+1} \left(
      \int\limits_{\tilde\mu_j}^{\tilde\mu^j_\mathrm{ex}}
      \frac{d\tilde\nu}{\sqrt{\Sigma_j + (-1)^j \tilde{\nu}^2}} -
      \int\limits^{\tilde\mu_{j+1}}_{\tilde\mu^j_{\rm ex}}
      \frac{d\tilde\nu}{\sqrt{\Sigma_j + (-1)^j \tilde\nu^2}}
    \right),\quad j=1,\dots,N/2
    .\label{alim}\\
    \beta a_c^{(N)} &=&
    \int\limits^{\tilde{\mu}_1}_{\tilde{\mu}_0}
    \frac{d\tilde\nu}{\sqrt{\tilde\nu^2-\tilde{\mu}_0^2}}.
    \label{endlim}
  \end{eqnarray}

For the sake of simplicity we introduce new variables $\xi_j={\tilde\mu_j}/{\tilde\mu_1}$, $\xi_1=1$, and
\begin{equation}
  \tilde \Sigma_j = \frac{\sqrt{(-1)^{j+1}\Sigma_j}}{\tilde\mu_1}
  = \sqrt{2(-1)^{j+1}\left(-\sum_{i=2}^{j}(-1)^i
  \xi_i^2+1-\frac{\xi_0^2}{2}\right)}
  . \label{Eq:tildeSigma2}
\end{equation}

The integration of Eq.~(\ref{alim}) separately for odd $j<N/2$, even $j<N/2$ and the integration of Eq.~(\ref{endlim}) give the system of equations for $a_c^{(N)}$ and miscellaneous variables $\xi_j$:
\begin{subequations}
  \begin{eqnarray}
    a_c^{(N)} = \pi -
    \arcsin\frac{\xi_{j+1}}{\tilde \Sigma_j} -
    \arcsin\frac{\xi_{j  }}{\tilde \Sigma_j}
    ,\mbox{ odd } j<\frac{N}{2} \label{Eq:a-for-odd-j}
    \\
    a_c^{(N)} = \ln
    \left[
      \frac{%
        \left(\xi_{j  } + \sqrt{\xi_{j  }^2-\tilde \Sigma_j^2}\right)
        \left(\xi_{j+1} + \sqrt{\xi_{j+1}^2-\tilde \Sigma_j^2}\right)
      }{\tilde\Sigma_j^2}
    \right]
    ,\mbox{ even } j<\frac{N}{2} \label{Eq:a-for-even-j}
    \\
    \beta a_c^{(N)}=\ln\left(\xi_0^{-1} + \sqrt{\xi_0^{-2}-1}\right)
    \label{betaac}
  \end{eqnarray}
  \label{Eq:ac-LargeBeta}
\end{subequations}

\emph{Even $N$, $\beta<1/2$, $\mu_c=\pi$}. Using Eq.~(\ref{Eq:mu-approx}) with $\mu_c=\pi$ and following the same procedure, we arrive to the following system of transcendental equations which define $a_c$:
\begin{subequations}
  \begin{eqnarray}
    \beta a_c^{(N)}=\frac{\pi}{2} - \arcsin \left(\frac{1}{\xi_0}\right)
    \\
    a_c^{(N)} = \pi -
    \arcsin\frac{\xi_{j+1}}{\tilde \Sigma_j} -
    \arcsin\frac{\xi_{j  }}{\tilde \Sigma_j}
    , \mbox{ for even } j < N/2 \label{Eq:a-for-odd-j2}
    \\
    a_c^{(N)} = \ln
    \left[
      \frac{%
        \left(\xi_{j  } + \sqrt{\xi_{j  }^2-\tilde \Sigma_j^2}\right)
        \left(\xi_{j+1} + \sqrt{\xi_{j+1}^2-\tilde \Sigma_j^2}\right)
      }{\tilde\Sigma_j^2}
    \right]
    ,\mbox{ for odd } j < N/2 \label{Eq:a-for-even-j2}
  \end{eqnarray}
  \label{Eq:ac-SmallBeta}
\end{subequations}
\end{widetext}

\begin{table}
  \begin{equation*}
    \begin{array}{|c|c|c|}
      \hline
      N & a_c^{N} & {\mbox{Solution of (\protect\ref{Eq:ac-LargeBeta})}} \cr
      \hline
      2 & 1.4639 & \xi_0=0.4392\cr
      4 & 1.1772 & \xi_0=0.5628, \xi_2=1.1469 \cr
      6 & 1.0060 & \xi_0 = 0.6451, \xi_2 = 1.1807, \xi_3 =
      1.3141\cr
      \hline
    \end{array}
  \end{equation*}
  \caption{
    The values of $a_c^{(N)}$ for $\beta=1$ (accuracy of calculation is $\pm 0.0001$).
  }
  \label{Tab:b1}
\end{table}
\begin{table}
  \begin{equation*}
    \begin{array}{|c|c|c|}
      \hline
      N & a_c^{N} & {\mbox{Solution of (\protect\ref{Eq:ac-SmallBeta})}}\cr
      \hline
      2 & 2.9277 & \xi_0=1.34429\cr
      4 & 1.2546 & \xi_0=1.0513, \xi_2=1.3734 \cr
      6 & 0.9833 & \xi_0 = 1.0310, \xi_2 = 1.2352, \xi_3 =
       1.3233\cr
      \hline
    \end{array}
  \end{equation*}
  \caption{
    The values of $a_c^{(N)}$ for $\beta=1/4$ (accuracy of calculation is $\pm 0.0001$).
  }
  \label{Tab:b4}
\end{table}

\begin{table}
  \begin{eqnarray}\nonumber
    \begin{array}{|c|c|c|}
      \hline
      N & a_c^{N} & {\mbox{Solution of Eq.~(\ref{Eq:ac-LargeBeta}) and (\ref{Eq:a-for-odd-j-bInf})}} \cr
      \hline
      2 & \pi/2 & {\mbox{---}}\cr
      4 & 1.3063 & \xi_2=1.2266\cr
      6 & 1.1343 & \xi_2 = 1.3290, \xi_3 = 1.6058 \cr
      8 & 1.0146  & \xi_2= 1.3772, \xi_3 = 1.7641, \xi_4 = 1.9065\cr
      \hline
    \end{array}
  \end{eqnarray}
  \caption{%
    The values of $a_c^{(N)}$ for $\beta=\infty$ (accuracy of calculation is $\pm 0.0001$)
  }
  \label{Tab:inf}
\end{table}

Now we consider infinitely long JJ. Similar to the finite LJJ, $a_c$ is not zero only for even $N$. For calculation of the crossover distance in this case we use equations (\ref{Eq:ac-LargeBeta}), with limit $\beta \to \infty$ and $\mu_0=0$. This modifies Eq.~(\ref{Eq:a-for-odd-j}) for $j=1$:
\begin{eqnarray}
  a_c^{(N)}=\frac{3}{4} \pi - \arcsin\frac{\xi_2}{\sqrt{2}}.
  \label{Eq:a-for-odd-j-bInf}
\end{eqnarray}
The rest of equations from the system (\ref{Eq:ac-LargeBeta}) stay the same.

To give an example, the values of $a_c^{(N)}$ calculated for $\beta=1$, $\beta=1/4$ and $\beta=\infty$ are presented in Tab.~\ref{Tab:b1}, Tab.~\ref{Tab:b4} and Tab.~\ref{Tab:inf}, respectively and are in accordance with Tab.~\ref{Tab:ac-instability}.
This technique of solving numerically a system of transcendental equations is rather effective and can be used to obtain the plots $a_c^{(N)}(\beta)$ shown in Fig.~\ref{Fig:a_c(b)}. Note, that these plots exactly coincide with the ones obtained in section \ref{Sec:FPS} using stability analysis for flat phase state.

The coincidence of the crossover distances obtained in two different ways implies that the transition between AFM state and flat phase state at $a=a_c$ happens because AFM
solution just cease to exist. It was believed before
\cite{Goldobin:SF-ReArrange} that transition takes place because one of
the states has lower energy. Now it is also clear why in Ref.~
\cite{Goldobin:SF-ReArrange} the hysteresis around $a_c$ was never seen!
Hysteresis usually takes place when one has two stable solutions having
different energies.

\begin{figure}[!htb]
  \begin{center}
    \resizebox{70mm}{!}{\includegraphics{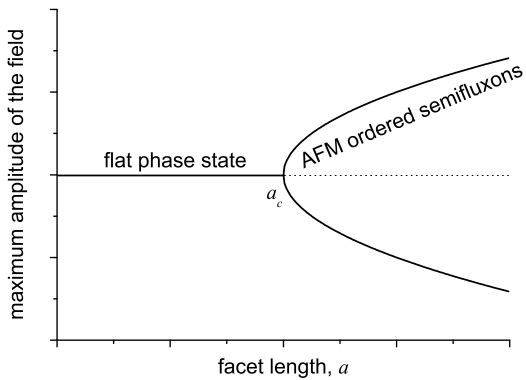}}
  \end{center}
  \caption{
    The sketch of the bifurcation diagram which show the transition from flat phase state to the state with AFM ordered semifluxon chain.
  }
  \label{Fig:Bifurcation}
\end{figure}

We can draw possible states of the system as a pitchfork bifurcation diagram shown schematically in Fig.~\ref{Fig:Bifurcation}. At small $a<a_c$ the flat state is the only solution and it is stable. At $a=a_c$ the flat phase solution looses its stability, and it is unstable at $a>a_c$ as indicated by the dotted line. In the same time, at $a=a_c$ two new solutions appear. Both correspond to AFM ordered chain of semifluxons but with different sign.

\section{Conclusions}
\label{Sec:Conclusion}

We have studied analytically the ground states in a 0-$\pi$-LJJ with
different number of facets of the length $a\sim\lambda_J$. We have shown that in the general case there is a crossover distance $a_c^{(N)}$ such that if the facet length $a<a_c^{(N)}$, the system is in the flat phase state ($\mu=const$) and contains no magnetic flux. In contrast, if $a>a_c^{(N)}$, the ground state consists of fractional vortices, each pinned at the phase discontinuity point. There may be more than one such state, especially for large $N$, but we focus our attention on the most natural one --- AFM ordered chain of semifluxons. The system chooses between flat phase state and AFM ordered chain of semifluxons not because of the energy competition as it was suggested earlier\cite{Goldobin:SF-ReArrange}, but because there is only one stable solution for given $a$, as shown in the bifurcation diagram Fig.~\ref{Fig:Bifurcation}: for $a<a_c^{(N)}$ AFM ordered semifluxon solution does not exist, while a flat phase state exists and is stable; for $a>a_c^{(N)}$ flat phase solution $\mu=const$ exists but is unstable, so the state is AFM ordered semifluxon chain. We have calculated the crossover distances $a_c^{(N)}$ and summarize our results as follows.
\begin{itemize}
  \item For odd $N$, $a_c=0$, semifluxons are always present.
  \item For even $N$ $a_c\ge 0$. The dependences of $a_c^{(2)}$, $a_c^{(4)}$ and $a_c^{(6)}$ on $b$ are shown in Fig.~\ref{Fig:a_c(b)}. In particular, for $b=a/2$ the $a_c^{(N)}=0$ and semifluxons are always present, for all other $b$, $a_c>0$.
\end{itemize}
Our calculations of $a_c$ agree with previous numerical and analytical results\cite{Goldobin:SF-ReArrange,Kato:1997:QuTunnel0pi0JJ,Kuklov:1995:Current0piLJJ}, but cover also the cases of larger $N$, arbitrary edge facets length $b$ and have much higher accuracy. We also show that in many cases the size of the edge facets $b$ can drastically affect the state of the whole system, especially when $b\approx a/2$ or $b\to0$ and $N=2$, as can be seen from Fig.~\ref{Fig:a_c(b)}.

We stress that we derived the position of bifurcation point $a_c$ approaching it from both flat phase state (from the left in Fig.~\ref{Fig:Bifurcation}) and from the state with AFM ordered chain of semifluxons (from the right in Fig.~\ref{Fig:Bifurcation}), and we got the same results. In the first approach, the system of equations [Eqs.~(\ref{Eq:a_c-instability1}) for $b>a/2$ or (\ref{Eq:a_c-instability2}) for $b<a/2$], that describe the (in)stability of the flat phase state, is particularly easy to solve numerically and the reader is encouraged to do so for his/her favorite values of $N$ and $b$ just setting proper seed value for $a_c$. Nevertheless, our derivation of more complex Eqs.~(\ref{Eq:ac-LargeBeta}) for $b>a/2$ and Eqs.~(\ref{Eq:ac-SmallBeta}) for $b<a/2$, which describes the disappearance of AFM ordered semifluxon chain and gives the same values of $a_c$, is not useless. This approach, although more complex, allows to find the existence region for more complex semifluxon states like \state{uudd}, which will be discussed elsewhere using the results obtained here.

We have also found that the crossover distance $a_c\propto1/\sqrt{N}$
for large $N$, see Eqs.~(\ref{Eq:a_c@LargeN-b0}) and (\ref{Eq:a_c@LargeN-b1}) or Eqs.~(\ref{Eq:int_approx}) and (\ref{Eq:g(beta)}) of appendix \ref{Sec:a_c@LargeN_app}. Having $a$ fixed, the longer 0-$\pi$-LJJ (larger $N$) favors configurations with semifluxons and, therefore, with magnetic flux, while shorter LJJ (smaller $N$) favors the state  without flux. Instead, if we fix the total LJJ length, the LJJ with smaller $a$ (large $N$) will favor flat phase state, while the LJJ with larger $a$ (small $N$) will favor the state with semifluxons.

In the future, it is quite interesting to consider the possibility to have less natural states like \state{uudd}. This will correspond to the additional branches on the bifurcation diagram and there will be clearly a minimum distance $a_c(\stat{uudd})>a_c(\stat{udud})$ ($a_c(\stat{udud})$ is the one found here) for which such a state is stable. For $a>a_c(\stat{uudd})$ there will be energy competition between various states, \eg, between \state{udud} and \state{uudd}. Next, in terms of studying classical and quantum tunneling between various states like \state{uudd} and \state{udud}, it is interesting to consider how $a_c$ depends on the applied magnetic field and bias current.

\begin{acknowledgments}
  This work was supported by the Deutsche Forschungsgemeinschaft
  (DFG) within project Si 704/2-1, and partially supported by the ESF programs ``Vortex'', ``Pi-shift'' and RFBR grants 03-01-06122 and 1716.2003.1.
\end{acknowledgments}

\appendix

\section{The only ``other'' value of $\mu_c=\pi/2$}
\label{Sec:mu_c=pi/2}

Now we show that if $a_c=0$, then  $\mu_c=\pi/2$. Let us subtract expression (\ref{mu}) for two end values of $\mu$ in the $n$-th odd interval: $\mu=\mu_n$ and $\mu=\mu_{n+1}$. For them  $x_{n+1}-x_{n }\approx \sqrt{\epsilon}$, $\alpha_{ n}\approx  \pm \sqrt{2}$, $x_{2n}^*$  are defined by the Eq. (\ref{xstar}). Having this, we receive:
\begin{eqnarray}
  \epsilon (\tilde\mu_{n+1}-\tilde\mu_{n}) &
  =&2\am\left[ (x_{n+1}-x_n^*)/\alpha_n,m_n \right]
  \nonumber\\
  &-& 2\am\left[ (x_n    -x_n^*)/\alpha_n,m_n \right] \approx
  \nonumber\\&
  \approx&\frac{2\sqrt{\epsilon}}{\alpha_n}\;
  \dn\left( \frac{x_{n}-x_n^*}{\alpha_n},\alpha_n^2 \right)
  + O(\epsilon).
  , \nonumber
\end{eqnarray}
where $\dn(x,m)=\sqrt{1-m\sin^2\am(x,m)}$ is the Jacobi elliptic function. There is only one term of the order $\sqrt{\epsilon}$ in this expression. Thus this term equals zero. Using expression (\ref{xstar}) for $x_n^*$ [$x_n^*=x_n -\alpha_n \F(\mu_n/2,m_n)$] with substitution $m_n\approx 2$  and $\partial\am(\xi,m)/\partial\xi=\dn(\xi,m)$ we get: $\dn[\F(\mu_c/2,2),2]=0$ which has the only solution inside of the interval $0<\mu<\pi$: $\mu_c=\pi/2$.

\section{Behavior of instability point at large $N$:
asymptotic relation for arbitrary $\beta$}
\label{Sec:a_c@LargeN_app}

The analysis in this section is based on the fact, that $(\beta_{j+2}-\beta_{j})/\beta_j\ll 1$ for $N\to\infty$. This allows us to approximate the function $\beta_j$ of discrete parameter, $j$, by the pair of continuous functions and derive the first order ordinary differential equation with boundary conditions for one of them. Solving it one gets the
implicit relation between $a_c$ and $N$. Without loss of
generality we assume that $N/4$ is odd in this section.

First, we consider the case $\beta>1/2$. Let us introduce two functions, corresponding to odd and even intervals: 
$A(n) = \beta_{2n-1}$, $B(n) = \beta_{2n}$, $n=1,2,\dots,N/4$ and rewrite the system (\ref{Eq:a_c-instability1}) in the following form
\begin{eqnarray}
  A(1)&=&-\arctan\left[ \tanh(\beta a_c) \right]
  ,\\
  \tan \left[ a_c+A(n) \right] &=& -\tanh\left[ B(n) \right]
  ,\label{Eq:a_c-instability1_app1}\\
  \tanh\left[ a_c+B(n) \right] &=& -\tan\left[ A({n+1}) \right]
  ,\label{Eq:a_c-instability1_app2}\\
  A({N/4-2})& \approx& A({N/4}) = -\frac{a_c}{2}.
  \label{Eq:a_c-instability1_app}
\end{eqnarray}
The index $n=1,2\dots,{N}/{4}-2$ in Eq.~(\ref{Eq:a_c-instability1_app1}) and $n=1,2\dots,{N}/{4}-1$ in Eq.~(\ref{Eq:a_c-instability1_app2}).

Now we may write
$A(n+1)\approx A(n) + A'(n)$, $A'(n)\ll A(n)$, where by definition
\[
  A'(n)=\lim_{\triangle n\to 0}
  \frac{A({n+\triangle n})-A(n)}{\triangle n}.
\]
Express $B(n)$ in terms of $A(n)$ using Eq.~(\ref{Eq:a_c-instability1_app1}) and expand the {\rhs} of the Eq.~(\ref{Eq:a_c-instability1_app2}) in series with respect to $A'$ keeping only linear term in $A'$. 
Then we have  (we now write "=" instead of "$\approx$"):
\begin{eqnarray}\label{Eq:beta_g_1_2}
  A'=f(A)=
  \frac{\cos A\left[ \sin a_c - \cos(a_c+2A) \tanh a_c \right]}%
       {\cos(a_c+A) - \sin(a_c+A) \tanh a_c},\\
  A(1)=-\arctan \left[ \tanh(\beta a_c) \right]
  ,\quad A(N/4)=-a_c/2.
\end{eqnarray}
where $A'$ is  positive, since $A(n+1)>A(n)$. The later is
consequence of the Eqs.~(\ref{Eq:a_c-instability1_app1}),
(\ref{Eq:a_c-instability1_app2}) and the property of the
increasing concaved down function $f$: $f(x_1+x_2)<f(x_1)+f(x_2)$. 
Thus the  above equation may be formally integrated:
\begin{eqnarray}\label{Eq:int}
  \int\limits_{A(1)}^{-a_c/2} \frac{\;d\beta}{f(\beta)} 
  = \left(\frac{N}{4}-3\right).
\end{eqnarray}
Although integration can be done explicitly, we stay with this
symbolic form for the sake of simplicity.

Similarly, for $\beta<1/2$ we get the same Eq.(\ref{Eq:int}) 
with different function $f$ and boundary condition $A(1)=-\atanh \left[ \tan(\beta a_c) \right]$:
\begin{eqnarray}\label{Eq:beta_l_1_2}
  f(A)=
  \frac{\cosh A\left[ \sinh a_c - \cosh(a_c+2 A) \tan a_c \right]}%
       {\cosh(a_c+A) + \sinh(a_c+A) \tan a_c}.
\end{eqnarray}
with negative $A'$.

Eq.(\ref{Eq:int}) gives us essentially function $N(a_c)$ rather
then desirable  $a_c(N)$. Fortunately $N(a_c)$ may be simply
inverted since  $a_c\ll 1$ everywhere in  these calculations. In fact, 
one can expand {\lhs} of (\ref{Eq:int}) in powers of $a_c$, keeping only the
leading term, which is of the order $a_c^{-2}$. Thus

\begin{equation}\label{Eq:int_approx}
  a_c \approx 2\sqrt{\frac{g(\beta)}{N}},
\end{equation}
where
%
\begin{subequations}
  \begin{eqnarray}
    g(\beta)&=&\sqrt{3} \arctan\left[\sqrt{3}(2\beta-1)\right]
    ,\; \beta>\frac{1}{2},\\
    g(\beta)&=&\sqrt{3} \arctan\left[\sqrt{3}(1-2\beta)\right]
    ,\;\beta<\frac{1}{2}.
  \end{eqnarray}
  \label{Eq:g(beta)}
\end{subequations}
In particular, for $\beta\to\infty$ we have $g= \pi\sqrt{3}/2$.

Thus for large $N$ and any length $b=\beta a$  of the end facets 
we have derived asymptotic relation $a_c \sim N^{-1/2}$,
which is in agreement with equations of the Sec.\ref{Sec:a_c@LargeN}.

\begin{figure}[!t]
  \begin{center}
    \resizebox{75mm}{!}{\includegraphics{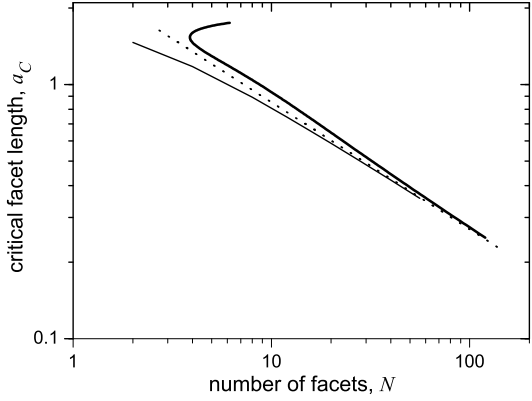}}
  \end{center}
  \caption{
    Asymptotic behavior of $a_c(N)$ for $\beta=1$.
    Solid bold line is received
    using Eq.~(\ref{Eq:int}); dotted  line corresponds to the
    Eq.~(\ref{Eq:int_approx}); solid thin line is result of
    numerical solution of the system (\ref{Eq:a_c-instability1}).
  }
  \label{Fig:Plots:ASSYMPTOTIC}
\end{figure}

Now we derive applicability condition for the equations of this
section. Eq.(\ref{Eq:int}) can be used for such $N$ that
\begin{eqnarray}
  \left|\frac{\max A'}{A}\right|\ll 1.
\end{eqnarray}
Investigation of the series in $a_c$ of the ratio $(\max A')/A$ shows that it does not have local extremum. Thus its maximum value can be taken only at the end points of the integration interval: $A=-a_c/2$ or $A=A(1)$. One can show that this point is $A(1)$ for all finite values of $\beta$ as well as for infinite $\beta$. 

Using leading (linear) term of the 
series of $A'$ in $a_c$ we get from Eqs.~(\ref{Eq:beta_g_1_2}) and (\ref{Eq:beta_l_1_2}):
\begin{eqnarray}
  A' &\approx& +2 a_c \sin ^2(A),\quad \beta>1/2,\\
  A' &\approx& -2 a_c \sinh^2(A),\quad \beta<1/2.
\end{eqnarray} 
Together with relation (\ref{Eq:int_approx}) and the fact that 
$A(1)\approx -\beta a_c$ for $\beta a_c \ll 1$, $-\pi/4 \lesssim A(1)<0$ 
for $\beta a_c \gtrsim 1$  and $A(1)=-\pi/4$ for
$\beta\to\infty$, we arrive to
\begin{eqnarray}
  N & \gg & 8 g ,\;\;\beta a_c\ll 1,\\
  N & \gg & 2\sqrt{2} g,\;\;\beta\to\infty \;\;{\mbox{or}}\;\;\beta
  a_c\gtrsim 1.
\end{eqnarray}

\bibliography{SF,QuComp,SFS,pi}

\end{document}